\documentclass[aps,prl,twocolumn,balance,superscriptaddress,floats,a4paper]{revtex4}
\usepackage{latexsym}
\usepackage{dcolumn}
\usepackage{graphicx}
\usepackage{amssymb}
\usepackage{graphics}
\usepackage{amsmath}
\usepackage{epsf}
\usepackage{float}
\usepackage{hyperref}
\hypersetup{colorlinks,linkcolor=blue,citecolor=blue,urlcolor=blue}
\usepackage[left=18mm,right=18mm,top=18mm,bottom=20mm]{geometry}
\usepackage{subfigure}

\begin{document}

\title{Equivalence of Effective Medium and Random Resistor Network models for disorder-induced unsaturating linear magnetoresistance}

\author{Navneeth Ramakrishnan}
\affiliation{Department of Physics and Center for Advanced 2D Materials, National University of Singapore, 117551, Singapore}

\author{Ying Tong Lai}
\affiliation{Yale-NUS College,  16 College Avenue West, 138527, Singapore}

\author{Silvia Lara}
\affiliation{Yale-NUS College,  16 College Avenue West, 138527, Singapore}

\author{Meera M. Parish}
\affiliation{School of Physics and Astronomy, Monash University, Victoria 3800, Australia}

\author{Shaffique Adam}
\affiliation{Department of Physics and Center for Advanced 2D Materials, National University of Singapore, 117551, Singapore}
\affiliation{Yale-NUS College,  16 College Avenue West, 138527, Singapore}

\date{\today}

\begin{abstract}
A linear unsaturating magnetoresistance at high perpendicular magnetic fields, together with a quadratic positive magnetoresistance at low fields, has been seen in many different experimental materials, ranging from silver chalcogenides and thin films of InSb to topological materials like graphene and Dirac semimetals.  In the literature, two very different theoretical approaches have been used to explain this classical magnetoresistance as a consequence of sample disorder.  The phenomenological Random Resistor Network model constructs a grid of four-terminal resistors, each with a varying random resistance. The Effective Medium Theory model imagines a smoothly varying disorder potential that causes a continuous variation of the local conductivity. Here, we demonstrate numerically that both models belong to the same universality class and that a restricted class of the Random Resistor Network is actually equivalent to the Effective Medium Theory. Both models are also in good agreement with experiments on a diverse range of materials.  Moreover, we show that in both cases, a single parameter, i.e.\ the ratio of the fluctuations in the carrier density to the average carrier density, completely determines the magnetoresistance profile.    
\end{abstract}

\maketitle

%
%\section{Introduction} 
%\label{sec:intro}
\textit{Introduction} -- The study of magnetoresistance (MR) is important both from a theoretical perspective, where it is a tool to probe the fundamental electronic properties of a material -- such as the band topology~\cite{kn:ashcroft1976} -- as well as from the point of view of applied physics -- for example, magnetic memory read-heads use MR as its working principle~\cite{hu2009linear,kn:nickel1995}.  There is a diverse range of mechanisms that can generate magnetoresistance including, for example, geometrical anisotropy~\cite{mcguire1975anisotropic,branford2005geometric}, multiple carrier channels~\cite{hwang2007transport}, and spin dependent scattering~\cite{baibich1988giant}.  At low temperatures, even weak magnetic fields disrupt the quantum interference between electron paths, which can give rise to either positive or negative magnetoresistance depending on the symmetry of the electron wavefunctions~\cite{hikami1980spin} (and this mechanism was recently used to probe the nature of spin-relaxation in monolayers of transition-metal dichalcogenides~\cite{schmidt2016quantum}).  In strong magnetic fields, the formation of Landau levels gives rise to Abrikosov quantum magnetoresistance, which shows a positive unsaturating linear MR~\cite{abrikosov1998quantum}, while a large negative magnetoresistance is often seen in ferromagnetic systems~\cite{matsukura1998transport}.  Longitudinal MR has also been suggested as a probe of the axial anomaly in Weyl semimetals~\cite{burkov2014chiral, goswami2015axial}.

In this Letter, we are interested in how macroscopic disorder can produce a magnetoresistance in the transverse configuration,
%concerned with transverse magnetoresistance, 
where the magnetic field is aligned perpendicular to the current. %and how it can occur due to macroscopic disorder.  
Here, the spatial variation of resistivity causes a component of the classical Lorentz force to have a component acting against the local direction of current flow. Disorder-induced MR gained attention about twenty years ago when it was invoked to explain the surprising experimental observation of unsaturating linear magnetoresistance in silver chalcogenides~\cite{xu1997large, husmann2002megagauss}.  The hallmark of disorder-induced MR is an unsaturating linear MR at high magnetic fields and a quadratic MR at low fields, that is relatively insensitive to temperature.  These features are now ubiquitous and have been observed in a wide class of materials including transition metal compounds~\cite{schnyders2015linear,von2009linear, xing2016large},
silver chalcogenides~\cite{hu2007nonsaturating}, monolayer graphene~\cite{ping2014disorder}, bilayer graphene~\cite{kisslinger2015linear, tiwari2009model}, Dirac semimetals~\cite{ramakrishnan2015transport,novak2015large,liang2015ultrahigh,wang2016large}, and black phosphorus~\cite{hou2016large}. There has also been significant theoretical interest in disorder induced MR in recent years~\cite{kozlova2012linear,song2015linear,alekseev2016magnetoresistance}. The fact that this MR persists to room temperature underscores its potential for technological impact~\cite{soh2002applied, xu1997large} -- this mechanism also provides a very tractable platform for technological applications where one simply needs to make a sample dirtier in order to enhance the magnitude of the MR. 

\begin{figure*}[!htb]
	\centering	
  \subfigure[]{\includegraphics[width=0.44\textwidth]{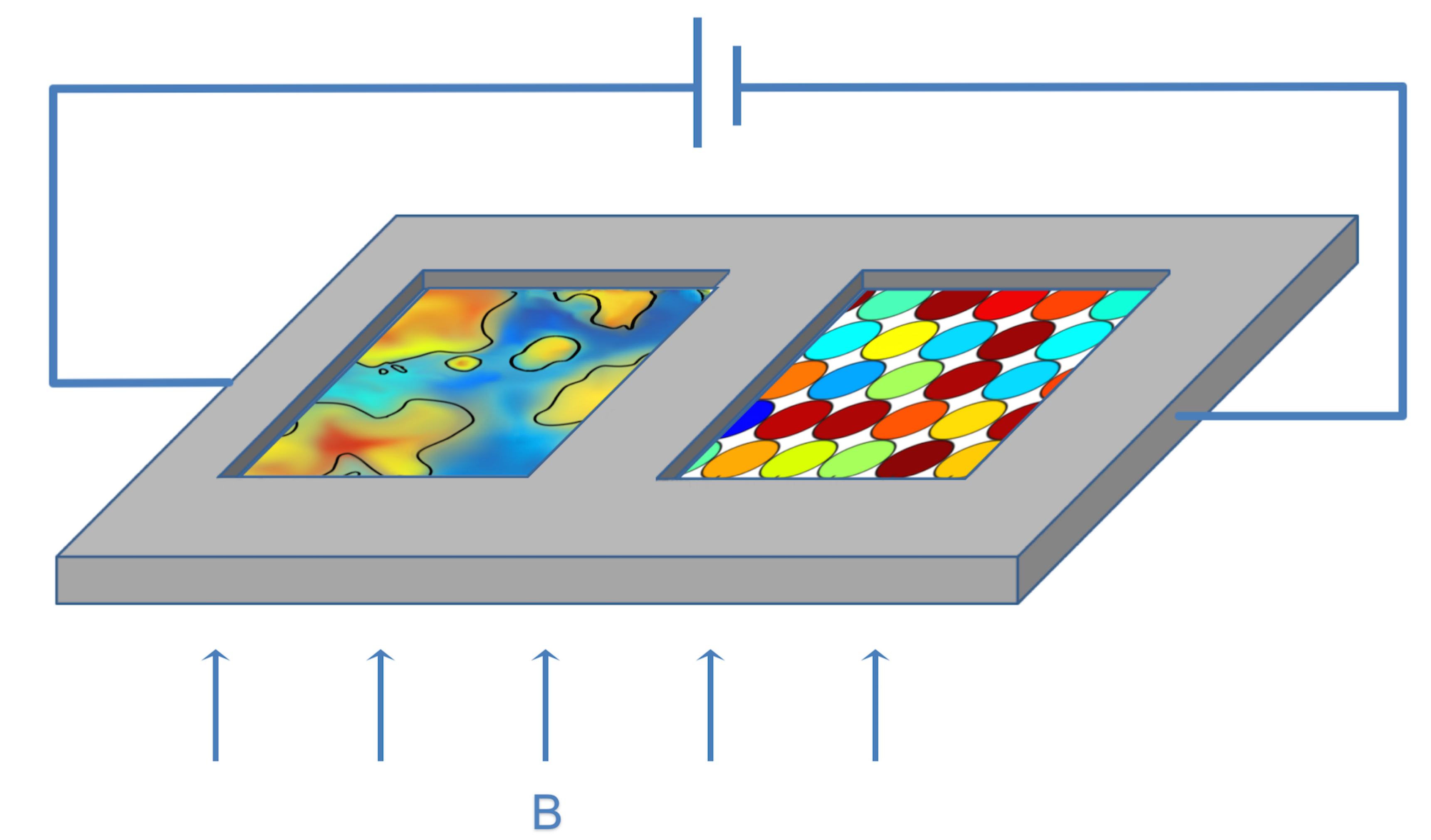}}
  \subfigure[]{\includegraphics[width=0.44\textwidth]{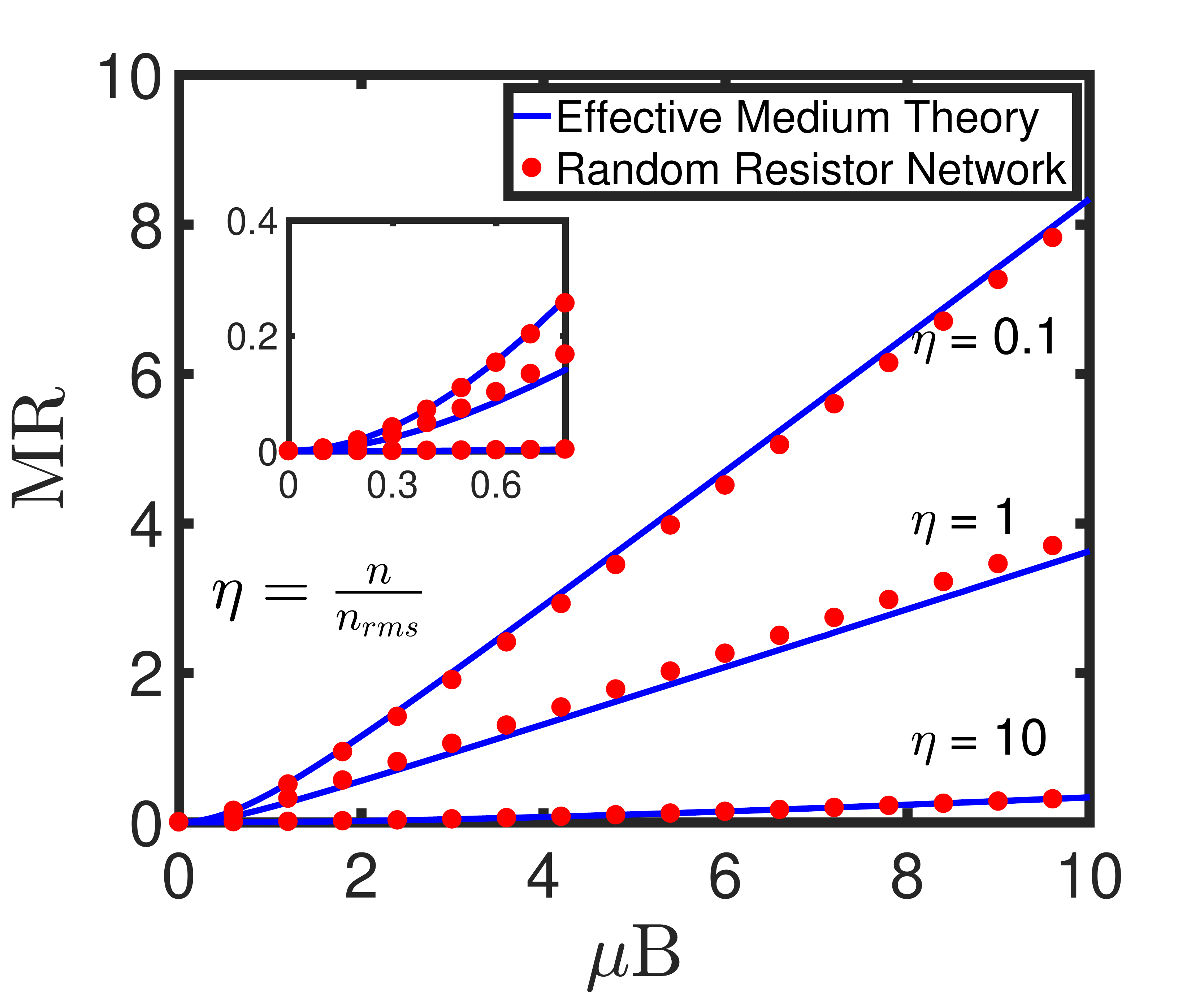}}
	\caption{(a) Disorder-induced magnetoresistance has been historically modeled by the Random Resistor Network (RRN) and the Effective Medium Theory (EMT), shown on the right and the left, respectively. The former characterizes the material as a network of four terminal resistors where the resistance value of each unit is random, while the latter models the material as an amalgamation of puddles where the conductivity is a continuous random variable (in this case $\sigma(n) = n\mu$, where $n$ is a random variable following a Gaussian distribution with mean $n_0$ and standard deviation $n_{\rm rms}$). (b) Equivalence of the RRN and EMT theories: As discussed in the main text, both the EMT and RRN theories can be formulated in terms of $n_0$ and $n_{\rm rms}$. The magnetoresistance is plotted here as a function of $\mu B$ for disorder parameter $\eta = \frac{n_0}{n_{\rm rms}} = 0.1, 1, 10$.  Notice that both theories depend only on the ratio of average carrier density to fluctuations in carrier density and give the same value of MR in both the low-field quadratic regime and the high-field linear regime, thus establishing the equivalence of the two theories.}
	\label{fig:cartoon}
\end{figure*}

The theoretical models of %developed to understand 
disorder induced MR build on ideas used to understand disordered media in the absence of a magnetic field, which itself has a long history~\cite{isichenko1992percolation}.  Two particular models have been especially successful, namely the random resistor network model (RRN)~\cite{kirkpatrick1973percolation} and the self-consistent effective medium theory~\cite{stroud1975generalized,landauer1978electrical}.  
For the case of binary mixtures in a cubic lattice, Kirkpatrick~\cite{kirkpatrick1971classical} showed almost fifty years ago that the two theories, although quite different in conception, actually gave the same result for the disorder-averaged resistance, provided that one stayed on the metallic side of the percolative metal-to-insulator transition.
%An important result, obtained almost fifty years ago by Kirkpatrick~\cite{kirkpatrick1971classical}, was that for the case of binary mixtures in a cubic lattice, provided that one stayed on the metallic side of the percolative metal-to-insulator transition, the two theories, although quite different in conception, actually gave the same result for the disorder-averaged resistance.  
However, when applied to the problem of magnetoresistance, the theoretical developments for these two approaches diverged again and the aforementioned experiments have been interpreted only in terms of one or the other.

More than a decade ago, one of us introduced~\cite{parish2003non} a Random Resistor Network model (RRN) to treat magnetoresistance in disordered silver chalcogenides using a network of four terminal resistors.   In addition to adding the effects of a magnetic field, 
our RRN had the value of each resistor drawn from a continuous probability distribution (unlike Ref.~\cite{kirkpatrick1971classical}) so that the magnetoresistance could depend on both the mean resistance and the fluctuations in resistance.  This model successfully captured the crossover from quadratic to linear MR and agreed with the magnitude of the observed magnetoresistance.  A few years later, Guttal and Stroud~\cite{guttal2005model} generalized the Effective Medium Theory (EMT) for a model of binary resistances to include a magnetic field, and also found the crossover from quadratic to linear MR.  More recently, while studying disordered monolayer graphene, two of us further developed~\cite{ping2014disorder} the EMT to include a continuous distribution of resistances and observed that the crossover from quadratic to linear MR depends only on a single parameter, namely the ratio of the fluctuations in carrier density to the average carrier density.  In particular, the magnitude of the MR did not depend on any of the properties of monolayer graphene such as the Dirac band-structure or the presence of both conduction and valence bands. 

While these two approaches for treating disorder-induced MR are very different, as illustrated in Fig.~\ref{fig:cartoon}a, our main result  is that the RRN and EMT are the same theory and can successfully model MR in a wide class of materials. We show in this work that the two models belong to the same universality class, and that the RRN can be configured such that there exists a one-to-one mapping with the EMT, as shown in Fig.~\ref{fig:cartoon}b. The two are equivalent such that for a given input disorder parameter, the resultant MR at any dimensionless magnetic field is the same for either theory.  

%This paper is organized as follows: in Section~\ref{sec:theory}, we recapitulate the two disorder models under consideration, and in Section~\ref{sec:comparison} we show clear evidence that the two models are equivalent. Finally, in Sec.~\ref{sec:experiments}, we compare the results of the unified model with experimental datasets from silver chalcogenides, graphene, Dirac semimetals and indium antimonide.

%\section{Theoretical background}
%\label{sec:theory}
%\subsection{Random resistor network (RRN)} 
%\label{subsec:RRN}
\textit{Random Resistor Network} -- The RRN model consists of a square grid of connected four terminal resistors. Two terminal resistors would not allow for transverse currents and hence, each resistor must have at least four terminals to capture the physics of magnetoresistance. Each resistor has an impedance matrix given by 

\begin{equation}\label{z_disk}
 z = \rho
     \left (
     \begin{array}{cccc}
      a & b & c & d \\  
      d & a & b & c \\ 
      c & d & a & b \\ 
      b & c & d & a
     \end{array}
     \right )
\end{equation}

%Integrate the $pi/4$ parameter into $\beta$ by adding it to $\mu_i$. Integrate the t parameter into n.
Here, $\rho$ is the disk scalar resistivity and we define a dimensionless magnetic field $\beta = \mu_0 B$, where $\mu_0$ is a phenomenological mobility-like parameter. The matrix elements depend on additional network specific parameters $\gamma$ and $\delta$, where we define $a=-\gamma+\beta$, $b=\gamma+\beta$, $c=\delta-\beta$ and $d=-\delta-\beta$.  

%
%Here, $\rho$ is the disk scalar resistivity and $t$ is the thickness of the resistor. We define the dimensionless magnetic field $\beta = \mu_i B$, where $\mu_i$ is a mobility like parameter. The matrix elements depend on additional network specific parameters such as $\gamma$ and $\delta$ as well as the dimensionless field $\beta$ and are given by: $a=-\gamma+\frac{\pi}{4}\beta$, $b=\gamma+\frac{\pi}{4}\beta$, $c=\delta-\frac{\pi}{4}\beta$ and $d=-\delta-\frac{\pi}{4}\beta$.  

The scalar resistivity value is given by $\rho = \frac{1}{n}$, where $n$ is a random variable chosen from a Gaussian distribution with mean $n_0$ and standard deviation $n_{\rm rms}$. (As seen in Fig.~\ref{fig:cartoon}b, we show that the magnetoresistance of the RRN only depends on $\eta = \frac{n_0}{n_{\rm rms}}$ and not on either $n_0$ or $n_{\rm rms}$ independently). The zero field mobility of the resistors is defined as $\mu = \frac{\langle\sigma\rangle}{n_0}$, where $\sigma$ is the conductivity in the absence of disorder and depends only on the parameters $\gamma$ and $\delta$. (Note that in the RRN model, the phenomenological $\mu_0$ defined above is an independent parameter in the theory from $\mu$, the mobility at zero magnetic field). 

In our work, we found it sufficient to use a 40$\times$40 square grid of random resistors.  Two possible sets of boundary conditions are possible - one is the hard boundary condition~\cite{parish2003non,parish2005}, where the left edge of the network is grounded and the right edge is set to a constant voltage, creating ``hard" edges. The other is periodic boundary conditions~\cite{hu2007nonsaturating}, where the voltage drop across every row of the network is kept constant, but no absolute voltage is specified along the edges. The input current on each row is also restricted to equal the output current, creating a ``periodic" tile in an infinite material. The resistivity of the network is solved using Kirchoff's laws for currents and voltages.  For the fixed boundary conditions, the MR is superlinear (sublinear) in the case of even (odd) network sizes, and in the limit of $N\rightarrow \infty$ one then observes the characteristic non-saturating quadratic-to-linear MR. One feature of the fixed boundary conditions is that it displays a finite MR even for the case of zero disorder, and the magnitude of the MR depends on the network size.  To avoid having this residual MR, for the rest of this work we focus only on the RRN with periodic boundary conditions. To summarize, the RRN takes as input parameters $\eta$ (i.e.\ the ratio of carrier density and density fluctuations), $\gamma$ and $\delta$ (network parameters from Eq.~\protect{\ref{z_disk}}), and the dimensionless magnetic field $\mu_0 B$, and gives the transverse magnetoresistance $\rho_{xx}$.  \\

% At first glance, since the RRN has additional input parameters $\gamma$ and $\delta$ that have no analogue in the EMT, one would not 
% ordinarily expect a mapping between the two theories.  However, we find that we can make the two theories identical by restricting the RRN
% in the following way. %makes the two theories identical.  
% First, we define the quadratic coefficient of magnetoresistance as $\mathcal{A} = \frac{MR}{(\mu B)^2}$ for $\mu B\ll 1$. For a given
% $\mu_0$, we choose any combination of $\gamma$ and $\delta$ (this choice is not unique) such that the quadratic coefficients in the EMT 
% and the RRN match for a single value of $\eta$ (e.g., $\eta = 0.1$).  However, with $\gamma$ and $\delta$ thus constrained, we find,
% surprisingly, that the MR in this restricted RRN depends only on two parameters $\eta$ and $\mu_0 B$.   

%\subsection{Effective Medium Theory (EMT)} 
%\label{subsec:EMT}
\textit{Effective Medium Theory} -- The Effective Medium Theory assumes that a sample is broken up into large macroscopic ``puddles" each with a given conductivity. One can understand the origin of these macroscopic puddles as a consequence of some inhomogeneous impurity potential. This spatially varying potential induces spatially varying local carrier densities and, consequently, varying local conductivities. Requiring that the average electric field over the sample be the same as the applied electric field, while allowing local variations in the electric field across different regions, one obtains the EMT equations. They are given in a general form below
\begin{equation} 
\label{eq:emt}
\int dn \, P[n, n_0, n_{\rm rms}] \frac{(\hat{\sigma}(n) - \hat{\sigma}^{E})}{\left(\mathbb{\hat{I}}_{2} + \frac{\mathbb{\hat{I}}_2}{d\hat{\sigma}_{xx}^{E}}(\hat{\sigma}(n) - \hat{\sigma}^{E})\right)} = 0,
\end{equation}
%\noindent 
where $d$ is the dimensionality of the system. $P[n, n_0, n_{\rm rms}]$ is the carrier distribution and is typically taken to be a Gaussian with mean $n_0$ and standard deviation $n_{\rm rms}$. $\hat{\sigma}(n)$ is the conductivity matrix as a function of the carrier density and $\hat{\sigma}^{E}$ is the effective conductivity that needs to be solved for. Note that the conductivity matrices include diagonal and off diagonal components when one considers the magnetic field,

\begin{equation}
\hat{\sigma}=\begin{pmatrix}
\sigma_{xx}& \sigma_{xy}\\
-\sigma_{xy}& \sigma_{xx}
\end{pmatrix} \, ,\ \,
\hat{\sigma}^{E}=\begin{pmatrix}
\sigma_{xx}^{E}& \sigma_{xy}^{E}\\
-\sigma_{xy}^{E}& \sigma_{xx}^{E},
\end{pmatrix}.
\end{equation}

With a transverse magnetic field, one obtains the form of the conductivity matrix elements as below
\begin{equation}
\sigma_{xx}=\sigma({n}) \frac{1}{1+\mu^2 B^2} \, ,\,
\sigma_{xy}=\sigma({n}) \frac{\mu B}{1+\mu^2 B^2},
\end{equation}
%\noindent 
where $\sigma(n) = n\mu$ is the zero field conductivity, $\mu$ is the mobility and $B$ is the applied magnetic field. The MR in the EMT only depends on $\eta$, the ratio of the mean of the Gaussian distribution and its standard deviation, and not on $n_0$ and $n_{\rm rms}$ independently.

\begin{figure}[!htb]
\includegraphics[angle=0,width=0.5\textwidth]{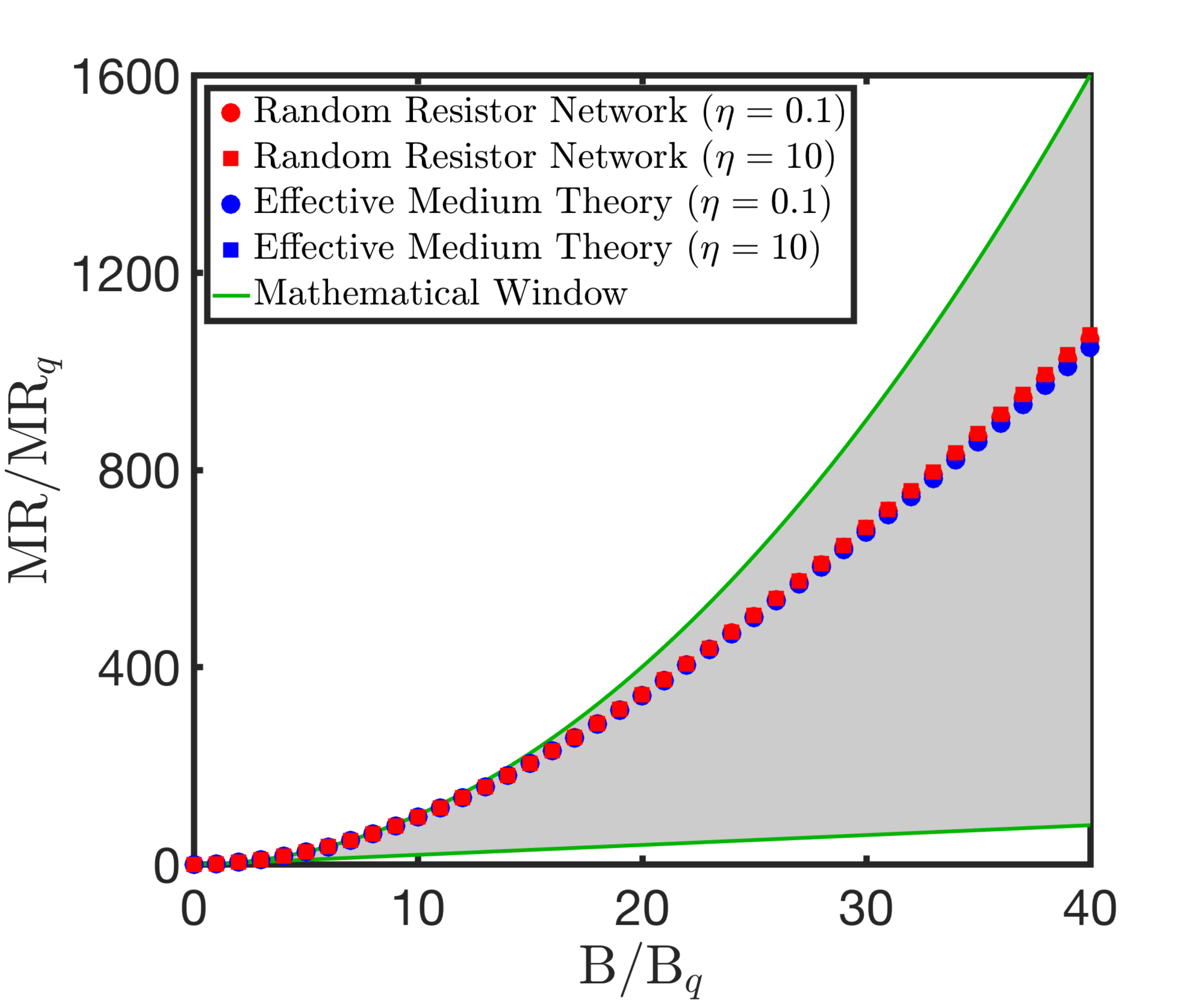}
\caption{\label{fig:theory1} (Color online) The numerical collapse of both the RRN and EMT disorder models onto a single curve when the $x$ and $y$ axes are scaled by $B_q$ and MR$(B_q)$ respectively shows that the theories belong to the same universality class. The mathematical window shows where all possible quadratic to linear curves may lie. Any analytic representation of these theories would only need two parameters corresponding to $B_q$ and MR$(B_q)$. Note that the disorder parameter $\eta = \frac{n_0}{n_{\rm rms}}$ covers a range spanning two orders of magnitude for both the EMT and the RRN.}
\end{figure}

\textit{Universality of unrestricted RRN and EMT} --  At first glance, since the RRN has additional input parameters $\gamma$ and $\delta$ that have no analogue in the EMT, one would not ordinarily expect any mapping between the two theories. However, both theories predict an MR that is quadratic at low fields and linear at high fields. We can therefore quantify the transition from quadratic to linear in order to identify if the two theories belong to the same universality class.  Defining $B_q$ as the magnitude of the magnetic field up to which the curve is quadratic (the exponent dropping below 1.999 is the numerical criterion used in this work), we scale the MR versus $B$ curves of both the EMT and the RRN by $B_q$ along the x-axis and $MR(B_q)$ along the y axis. Next, we construct a mathematical window for all possible crossovers from quadratic at low fields to linear at high fields.  The family of all quadratic to linear functions that have been scaled this way lie inside the window shown in Fig.~\ref{fig:theory1} that is bounded above by the curve $y = x^2$ and bounded below by the curve $y = x^2$ for $0\leq x\leq 1$ and $y = 2x - 1$ for $x>1$. The upper bound represents an infinitely slow transition from quadratic to linear after $B_q$, while the lower bound represents an instantaneous transition at $B_q$. Remarkably, both the EMT and the RRN curves show a numerical collapse onto the same crossover curve for all values of $\eta$. This collapse is valid even when the network parameters $\gamma$ and $\delta$ in the RRN are chosen arbitrarily. The numerical collapse highlights that the EMT and the unrestricted RRN belong to the same universality class.  In both cases, the full theory can always be reconstructed from the universal scaled theory with only two parameters corresponding to $B_q$ and MR($B_q$).  Experimental data could be scaled in the same way to determine if the MR mechanism also belongs to this universality class.

\begin{figure}[!htb]
\includegraphics[angle=0,width=0.5\textwidth]{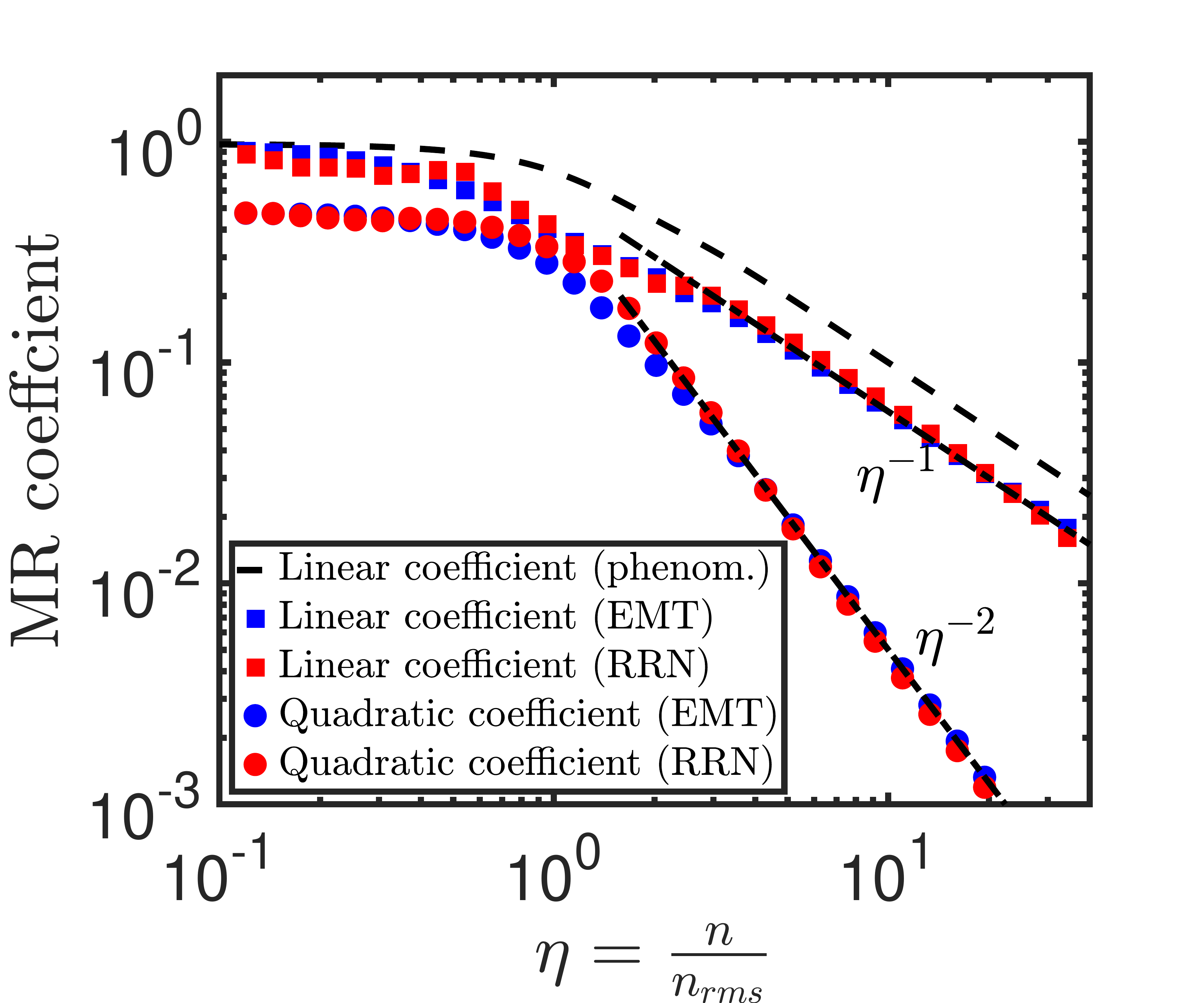}
\caption{\label{fig:coeffcompare} (Color online) The quadratic coefficient of magnetoresistance $\mathcal{A} = \frac{MR}{(\mu B)^2}$ in the limit $\mu B\ll 1$ and the linear coefficient of MR defined as $\mathcal{B} = \frac{\Delta MR}{\mu \Delta B}$ in the limit $\mu B \gg 1$ for both theories. Agreement of $\mathcal{A}$ at $\eta = 0.1$ is through fitting, but agreement at all other $\eta$ for $\mathcal{A}$ and $\mathcal{B}$ shows clear evidence that the two theories are equivalent. This is in contrast to the the quadratic to linear phenomenological formula that has been previously used to characterize MR~\cite{cho2008charge}, which shows quantitatively different results for the relationship between $\mathcal{A}$ and $\mathcal{B}$. Note that at large $\eta$, we have $\mathcal{A}\sim\eta^{-2}$ and $\mathcal{B}\sim\eta^{-1}$.}
\end{figure}

\textit{Equivalence of restricted RRN and EMT} -- The additional parameters $\gamma$ and $\delta$ in the RRN may be restricted to allow us to make a more rigourous comparison between the two models.   First, we define the quadratic coefficient of magnetoresistance as $\mathcal{A} = \frac{MR}{(\mu B)^2}$ for $\mu B\ll 1$. For a given $\mu_0$, we choose any combination of $\gamma$ and $\delta$ (this choice is not unique) such that the quadratic coefficients in the EMT and the RRN match for a single value of $\eta$ (e.g.\ $\eta = 0.1$).  With $\gamma$ and $\delta$ thus constrained, we find that the MR in this restricted RRN depends only on two parameters $\eta$ and $\mu_0 B$. Thus, both the EMT and the RRN take three input parameters ($\eta$, $\mu$ and $B$ for the EMT and $\eta$, $\mu_0$ and $B$ for the RRN) and output the resistivity. Moreover, as shown in Fig.~\ref{fig:cartoon}b, the two theories produce identical MR curves for all values of $\eta$.

Having established the equivalence of the two theories, we now examine some of the features of this unified theory. Figure~\ref{fig:coeffcompare} shows the quadratic and linear coefficients of MR ($\mathcal{A}$ and $\mathcal{B}$) for the EMT and the RRN over two orders of $\eta$. While agreement of $\mathcal{A}$ at $\eta=0.1$ is by choice of $\gamma$ and $\delta$, the agreement at all other $\eta$ for both $\mathcal{A}$ and $\mathcal{B}$ provides convincing evidence that two theories are indeed equivalent. Moreover, we find that for large $\eta$, i.e. in more homogeneous systems, the magnetoresistance persists and the quadratic coefficient goes as $\mathcal{A}\sim \eta^{-2}$ while the linear coefficient goes as $\mathcal{B}\sim \eta^{-1}$. This is in 
agreement with the existing results in the literature~\cite{ping2014disorder}, where the quadratic coefficient of the EMT was indeed found to obey this relationship.  

We remark that not all quadratic to linear functions, such as the phenomenological quadratic-to-linear formula of Ref.~\cite{cho2008charge}, agree with the EMT and the RRN. In that formula, one had $MR = \sqrt{1 + 2\mathcal{A}(\mu B)^2} - 1$,
which implies $\mathcal{B} = \sqrt{2\mathcal{A}}$. Despite obeying the same qualitative relationship between $\mathcal{A}$ and
$\mathcal{B}$, the magnitude of the coefficient $\mathcal{B}$ is different compared to the EMT and RRN for a given $\mathcal{A}$, as shown in Figure~\ref{fig:coeffcompare}.

%
%\section{Comparison with experiments} 
%\label{sec:experiments}

\begin{figure}[!htb]
\includegraphics[angle=0,width=0.5\textwidth]{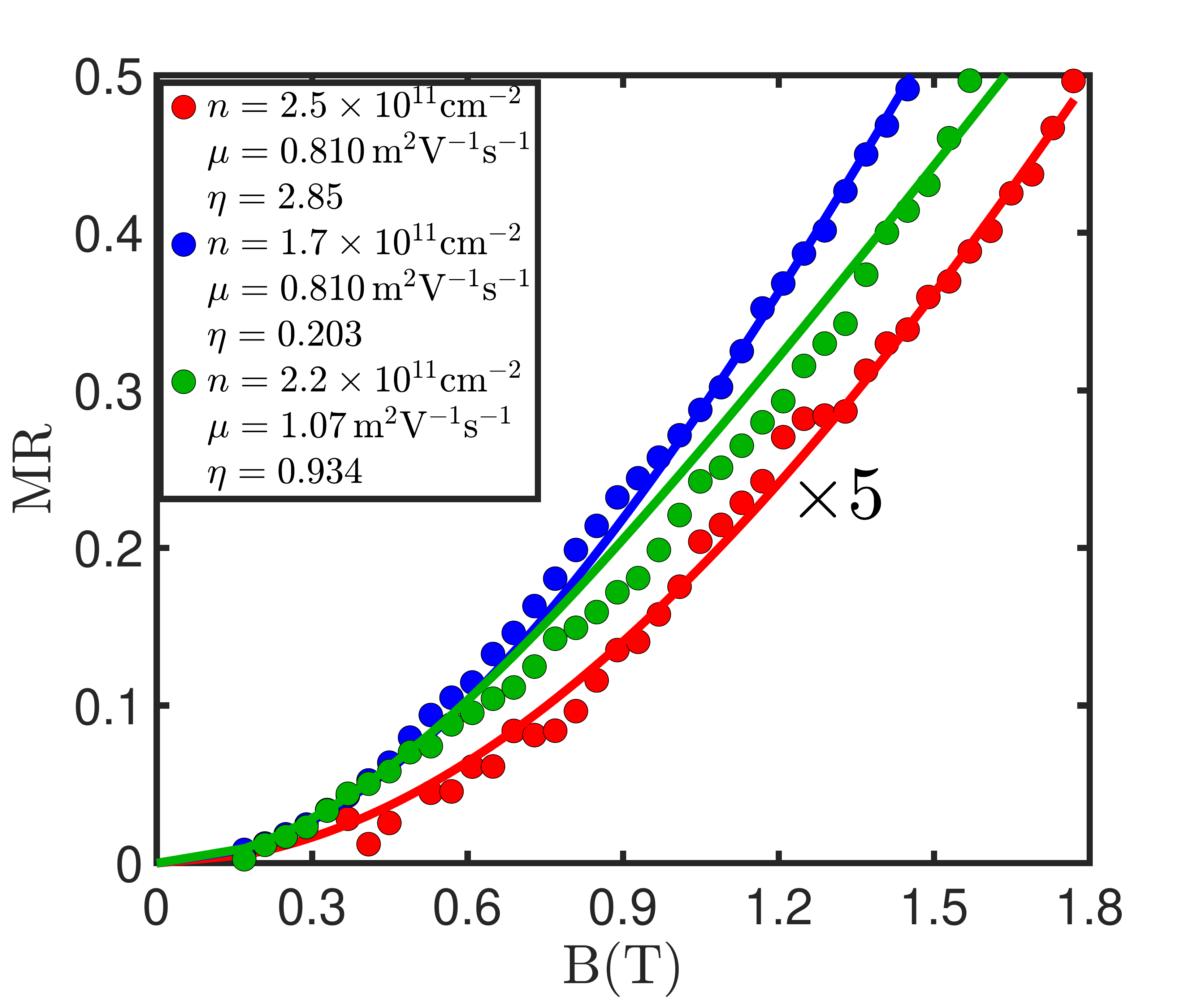}
\caption{\label{fig:exptgraphene} (Color online) Magnetoresistance data from the graphene samples of Ref.~\cite{ping2014disorder} show good agreement with the EMT-RRN formalism. In these experiments, the mobility is known (one measures the mobility in the regime with low disorder by using a back gate to increase carrier density) and $\eta$ is a fit parameter. The magnetoresistance data can be used as a measure of disorder, $\eta$, or vice versa.}
\end{figure}

\textit{Comparison with experiments} -- Next, we compare the EMT-RRN theory with MR experiments from several materials. We begin with the graphene samples from Ref.~\cite{ping2014disorder}. In this case, the mobility of the samples is determined through a Hall measurement at high carrier densities. Subsequently, a back gate is used to lower the carrier density and investigate the disordered regime. Figure~\ref{fig:exptgraphene} shows three MR curves for different samples, all of which can be fitted using the EMT-RRN formalism. One can calculate the disorder parameter $\eta$ through these fits and this can be independently verified through STM measurements. Alternatively, knowing $\eta$, one can make quantitative predictions regarding the MR. 

\begin{figure*}[!htb]
\centering
  \subfigure[]{\includegraphics[width=0.31\textwidth]{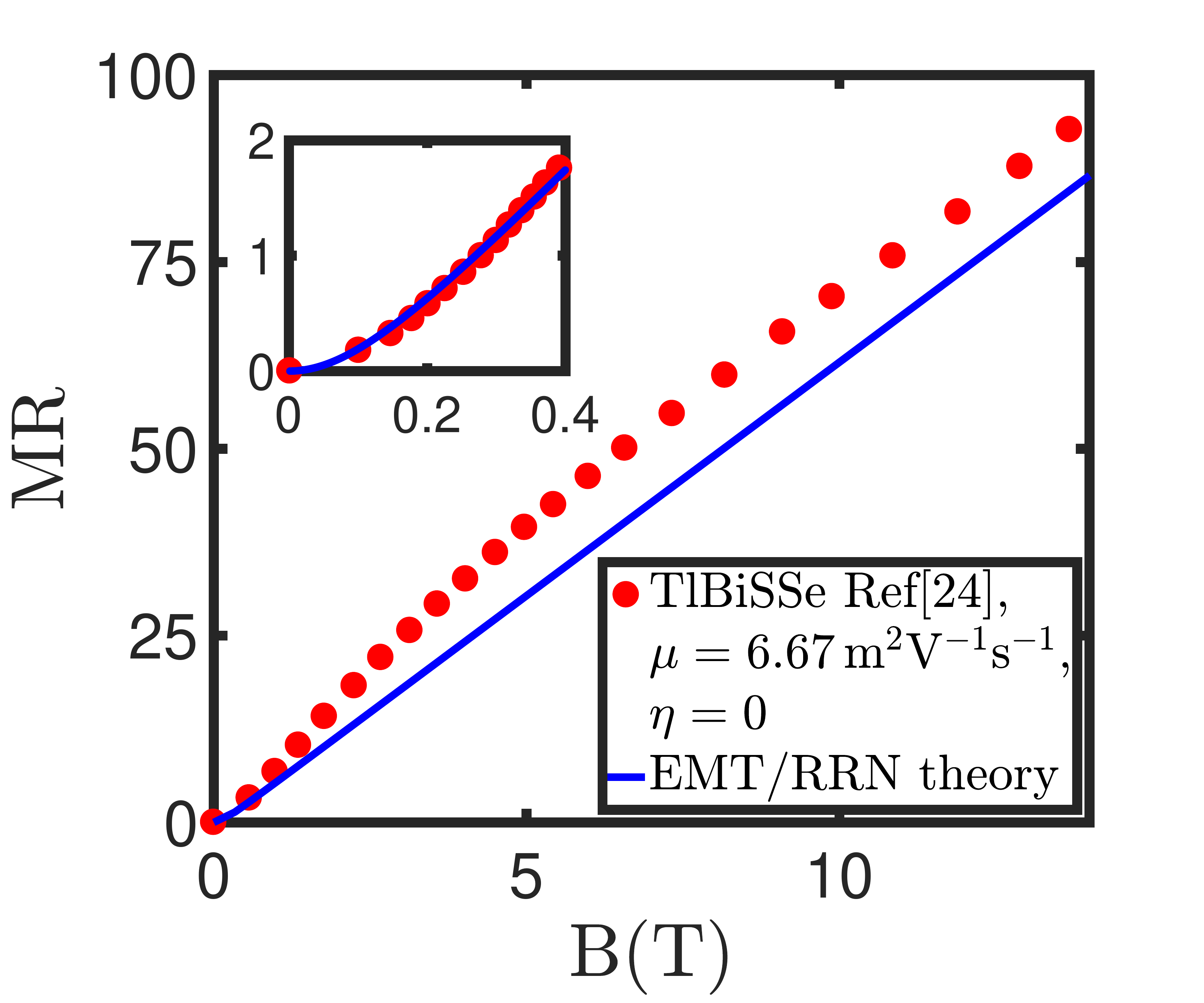}\label{fig:expt1}} 
  \subfigure[]{\includegraphics[width=0.31\textwidth]{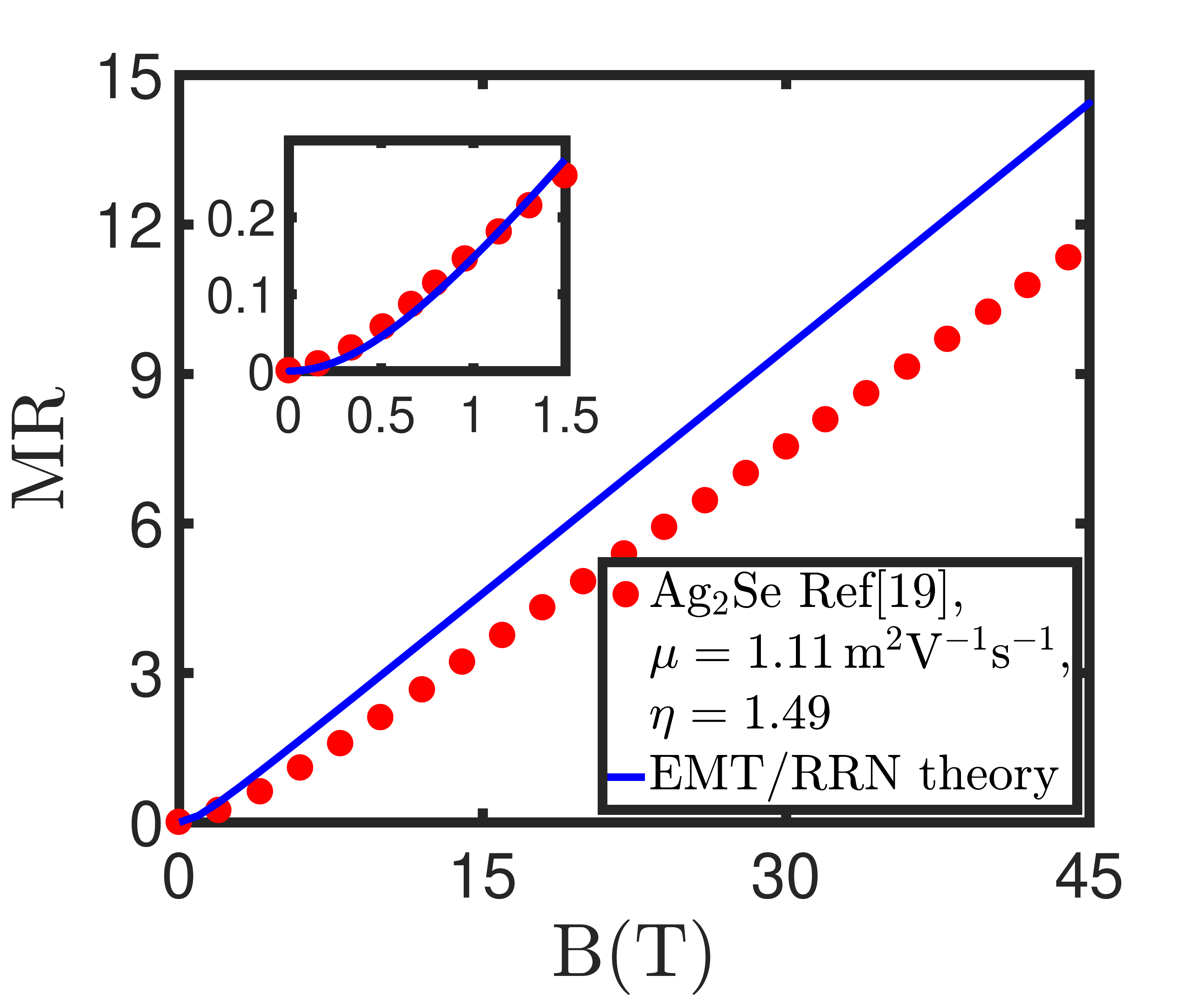}\label{fig:expt2}} 
  \subfigure[]{\includegraphics[width=0.31\textwidth]{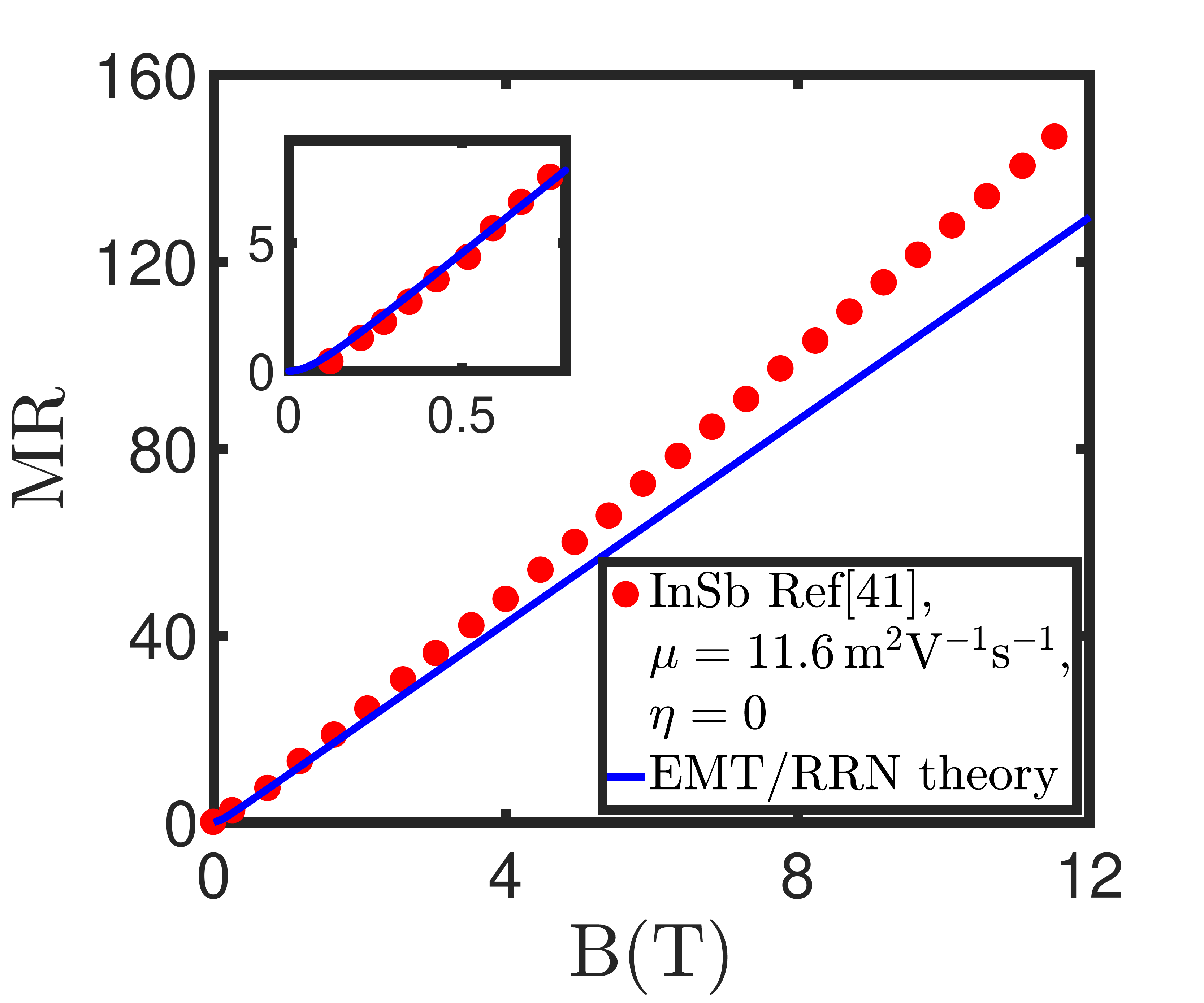}\label{fig:expt3}}
  \caption{(Color online) Experimental magnetoresistance data using Dirac semimetal TlBiSSe (Ref.~\cite{novak2015large}), silver selenide (Ref.~\cite{hu2007nonsaturating}) and indium antimonide (Ref.~\cite{hu2008classical}) are compared against the EMT-RRN formalism. A back gate to increase the carrier density and measure mobility was not possible and hence there are two free parameters, $\eta$ and $\mu$, for the theory. We therefore only use the low field data (shown in the insets) to find the best fit for both $\eta$ and $\mu$. Good agreement is found in each case in the high field linear regime and the fitted $\mu$ is within the range of the experimental Hall mobilities of these materials. We conclude that the disorder induced magnetoresistance in these diverse range of materials can indeed be explained by this formalism.}
  \label{fig:expt}
\end{figure*}

We also compare the theory against MR experiments performed on InSb, Dirac semimetal TlBiSSe and Ag$_2$Se. In these experiments, MR data is available for a low field regime and a high field regime. We use the low field data to fit both $\mu$ and $\eta$ and check the agreement of the theory (which has no more free parameters) and 
experiment in the high field regime. Typically, the low field regime lasts until around 1T and the error in the MR corresponding to the highest magnetic field value is under 20\% for each of the three experiments.

Strictly speaking, one should not compare the mobility values obtained through the theory with mobility values obtained through a Hall measurement. Hall measurement mobilities are only valid at high $\eta$ and these experiments did not have a back gate to increase the carrier density and then do the Hall measurement. However, as a rough guide we may still use them for comparison, and we find that the fitted values of $\mu$ are in good agreement with the reported Hall measurement mobility values for comparable samples. The reported Hall mobility values for InSb~\citep{rode1971electron, hu2008classical}, Dirac semimetal TlBiSSe~\cite{novak2015large} and Ag$_2$Se~\cite{ferhat2000thermoelectric} are 9.4 m$^2$V$^{-1}$s$^{-1}$, 5 m$^2$V$^{-1}$s$^{-1}$ and 0.3 - 1.5 m$^2$V$^{-1}$s$^{-1}$ respectively. The predictions based on fitting the curves with the EMT-RRN theory yield mobility values of 11.6 m$^2$V$^{-1}$s$^{-1}$, 6.7 m$^2$V$^{-1}$s$^{-1}$ and 1.1 m$^2$V$^{-1}$s$^{-1}$ respectively. 

The agreement between the experimental MR curves and the EMT-RRN theory for such a diverse class of experiments shows that this mechanism for disorder induced magnetoresistance is widespread and provides good validation of our theory.  Since the spatial density profile in 2D materials can be measured with local probe methods like STM, %can be done for 2D materials, 
such a measurement, in combination with magnetotransport data, would provide a rigorous test of this theory.

%\section{Acknowledgements}
\textit{Acknowledgements} -- This work was supported by the National Research Foundation of Singapore under its fellowship program (NRF-NRFF2012-01)  and by the Singapore Ministry of Education and Yale-NUS College through Grant No.~R-607-265-01312. This research was supported by the ARC Centre of Excellence in Future Low-Energy Electronics Technologies (Project Number CE170100039). We thank Michael Fuhrer, Derek Ho, Cameron Love, Alex Rodin and Indra Yudhishtira for useful discussions.  The EMT code written in MATLAB can be found at https://github.com/silvia-lara/effective-medium-theory.  LYT and SL contributed equally to this work. 

\bibliography{RRNbib}

\end{document}